\RequirePackage{lineno}
\documentclass[10pt,numberedappendix]{emulateapj}

\usepackage{graphicx}
\usepackage{bm}
\usepackage{amssymb,amsmath}
\usepackage{mathrsfs}
\usepackage{latexsym}
\usepackage{color}
\usepackage[normalem]{ulem} 
\usepackage{dcolumn}
\usepackage[colorlinks=true,citecolor=blue,urlcolor=blue]{hyperref}
\usepackage[usenames,dvipsnames]{xcolor}

\allowdisplaybreaks

\newcommand{\cg}[1]{\textcolor{black}{ #1}}

\newcommand{\CCA}{\affiliation{Center for Computational Astrophysics, Flatiron Institute, 162 5th Ave, New York, NY 10010}}
\newcommand{\BHI}{\affiliation{Black Hole Initiative, Harvard University, 20 Garden St., Cambridge, MA 02138}}

\definecolor{azgreen}{rgb}{0.03,0.47,0.19}
\definecolor{kcmagenta}{rgb}{0.54, 0.17, 0.88}
\definecolor{chorange}{rgb}{0.851, 0.372, 0.007}

\begin{document}

\title{Distinguishing binary neutron star from neutron star-black hole mergers with gravitational waves}

\author{Hsin-Yu Chen}\BHI
\author{Katerina Chatziioannou}\CCA

\date{\today}

\begin{abstract}

The gravitational-wave signal from the merger of two neutron stars cannot be easily differentiated from 
the signal produced by a comparable-mass mixed binary of a neutron star and a black hole. 
Indeed, both binary types can account for the gravitational-wave signal GW170817 
even if its electromagnetic counterpart emission is taken into account.
We propose a method \cg{that} {requires neither information from the post-inspiral phase of the binary nor an electromagnetic counterpart} to identify mixed binaries of neutron stars merging with low-mass black holes 
using gravitational-waves alone. 
This method is based on the fact that certain neutron star properties that can be measured with gravitational-waves
are common or similar for all neutron stars.
For example all neutron stars share the same equation of state and if the latter is hadronic, neutron stars have similar radii.
If a mixed binary is misidentified as a neutron star binary, the inferred neutron star 
properties will be misestimated and appear as outliers in a population of low-mass binaries.
We show that as few as {$\sim 5$} low-mass events will allow for the identification of the type of one 
event at the $80\%$ confidence level. 
We model the population of low-mass binaries with a hierarchical mixture model
and show that we can constrain the existence of mixed binaries or 
measure their abundance relative to neutron star binaries to 
$\sim 0.1$ at the $68\%$ credible level with 100 events. 

\end{abstract}

\maketitle
  
\section{Introduction}

The gravitational-wave (GW) event GW170817 detected by Advanced LIGO~\citep{TheLIGOScientific:2014jea} and Virgo~\citep{TheVirgo:2014hva}
 is consistent with the merger of two neutron stars (BNS)~\citep{TheLIGOScientific:2017qsa}. 
 Although the GW data place a lower limit on the compactness 
of the two coalescing bodies, objects more compact than neutron stars (NSs) are not ruled out~\citep{Abbott:2018wiz}. 
Arriving after the GW signal, the electromagnetic (EM) counterparts {GRB 170817A~\citep{Monitor:2017mdv,2017ApJ...848L..14G,2017ApJ...848L..15S}} 
and kilonova AT 2017gfo, e.g. \cite{2017Sci...358.1556C,Soares-Santos:2017lru},
imply the {presence} of at least one NS in the binary. 
However, we still can not exclude the possibility of GW170817 {being} a merger of a NS and a black hole (NSBH)~\citep{Abbott:2018wiz,Hinderer:2018pei,Coughlin:2019kqf}.


{X-ray binaries suggest a lack of BHs with mass below $5M_{\odot}$~\citep{1998ApJ...499..367B,2011ApJ...741..103F,2012ApJ...757...36K}, 
but the origin of this mass gap between BHs and NSs is not fully understood~\citep{Kreidberg_2012,2012ApJ...757...91B}. 
Recently, the discovery of a $3.3^{+2.8}_{-0.7}M_{\odot}$ unseen companion of the giant star 2MASS J05215658+4359220 
further challenged the existence of the mass gap~\citep{2019Sci...366..637T}.
Scenarios for the production of low-mass BHs 
include primordial density fluctuations~\citep{Carr:2016drx}, slow supernova explosions~\citep{2012ApJ...757...91B}, 
 mergers of NSs~\citep{Faber2012}, and interactions of dark matter and NSs~\citep{Bramante:2017ulk}.
 Low-mass binary mergers can potentially help 
study the black hole mass distribution~\citep{LIGOScientific:2018jsj}, but 
probing the existence of objects in the mass gap is challenging~\citep{Littenberg_2015,10.1093/mnrasl/slv054}.}

As already noted in~\cite{TheLIGOScientific:2017qsa}, though, constraining the component 
masses in $\sim (0.5-2) M_{\odot}$ does not 
definitively prove the type of the binary. 
For that we also need to detect (or rule out) tidal interactions in the binary with GWs,
 quantified through the NS tidal deformabilities~\citep{Flanagan:2007ix,Hinderer:2007mb}.
{For binary mergers, the} individual tidal parameter of each star is difficult to measure;
instead constraints are placed on a combination of masses and tidal deformabilities, 
$\tilde{\Lambda}$~\citep{Wade:2014vqa}. For an NSBH that is not particularly loud (signal-to-noise 
ratio less than $\simeq 30$), the tidal deformability is generally difficult to 
measure\cg{~\citep{2011PhRvD..84j4017P,2012PhRvD..85d4061L,2014PhRvD..89d3009L,2017PhRvD..95d4039K,2020arXiv200208383T}. }

GW170817 data place a lower limit on $\tilde{\Lambda}$ subject to the assumption of 
small spins~\citep{Abbott:2018wiz}; the data are nonetheless consistent with 
a highly spinning BH binary. At the same time, a nonzero $\tilde{\Lambda}$ only suggests the presence of
\emph{one} NS, still allowing for the NSBH scenario. 
Further analysis of the EM counterpart remains inconclusive and cannot rule out the NSBH scenario~\citep{Hinderer:2018pei,Coughlin:2019kqf}.
Similar analyses for near-future detections are subject to the availability and interpretation of an EM counterpart,
while post-merger information~\citep{Abbott_2017,Chatziioannou:2017ixj,Torres-Rivas:2018svp} or evidence for disruption~\citep{Pannarale:2015jia}
will likely be buried in detector noise.

The misidentification of a low-mass NSBH for a BNS
can have dire consequences for our ability to accurately measure the radius of NSs with GWs. Indeed
{a GW} analysis of a NSBH assuming it is a BNS underestimates the true radius~\citep{Yang:2017gfb}. 
The amount of bias depends on the mass of the BH as the tidal deformability
is a steeply decreasing function of the mass. Misidentifying a $\sim 2 M_{\odot}$ BH
for a NS induces a negligible error, while misidentifying a $\sim 1 M_{\odot}$ BH
can lead to a radius error of multiple km.

We present a method to distinguish between BNSs and low-mass NSBHs using their
GW signals alone. We take advantage of the inferred radius bias that is incurred for NSBHs and the
fact that \cg{the} {NS radius depends weekly on their mass} for hadronic equations of state (EoSs). 
A population of low-mass binaries of mostly BNSs and a few NSBHs will 
lead to inferred radii that are either approximately common (the BNSs) or outliers
(the NSBHs).
We show that BNSs and NSBHs can be identified within such a mixed population based on 
their inferred radii with high confidence, allowing us to
estimate the rate of low-mass NSBHs and achieve 
an unbiased measurement of NS radii.

\section{Method and Results}

Consider a low-mass binary with estimated component masses in the range $(0.5,2)M_{\odot}$, consistent 
with known NS masses and GW170817~\citep{TheLIGOScientific:2017qsa,Abbott:2018wiz,Abbott:2018exr}.
 In this mass range and for hadronic 
 EoSs that can support at least $2M_{\odot}$ NSs~\citep{Antoniadis:2013pzd}, 
the NS radius is expected to be constant to within a few hundred meters~\citep{Ozel:2016oaf}. 
If the system is a BNS, then we can infer this almost-common radius, but for a misidentified
 NSBH any radius estimate will be biased. 

To quantify the bias we assume that the first binary component is a NS (the
presence of which can be confirmed by detection of an EM counterpart or tidal effects) with mass $m_1$ and  tidal deformability $\Lambda_1$, 
while the second component could be either a NS or a BH with mass $m_2$ and 
tidal deformability $\Lambda_2$ ($\Lambda_2=0$ for BHs).
In either scenario, the leading order tidal effects will be
encoded in the GW phase through
\begin{align}
\tilde{\Lambda} &\equiv \frac{16}{13}\frac{(m_1+12m_2)m_1^4 \Lambda_1 +(m_2+12m_1)m_2^4 \Lambda_2  }{(m_1+m_2)^5}.
\end{align}
A GW analysis estimates $\tilde{\Lambda}_{\textrm{est}}=\tilde{\Lambda}$ if the source is a BNS, 
or $\tilde{\Lambda}_{\textrm{est}}=\tilde{\Lambda}(\Lambda_2=0)$ if it is a NSBH and an error.

The NS radius is then inferred from $\tilde{\Lambda}_{\textrm{est}}$ with use of two relations that do 
not sensitively depend on the EoS. 
The first relates the NS compactness to the
tidal deformability $C=C(\Lambda)$, and can be used to obtain the radius from the tidal 
deformability and the mass, $R=m/C(\Lambda)$~\citep{Maselli:2013mva,Yagi:2016bkt}. 
This relation holds for any NS, regardless of whether it is part of a NSBH or a BNS. 
{The second relation applies to BNSs only and it relates the individual tidal 
deformabilities of the two binary components given their 
mass ratio~\citep{Yagi:2015pkc,Chatziioannou:2018vzf}.}
 
Working under the assumption that the binary is a BNS (a common assumption for GW170817), we
use the two EoS-insensitive relations to obtain $R_{\textrm{\mbox{\tiny BNS}}}$ ($R_{\textrm{\mbox{\tiny NSBH}}}$), 
the radius estimate if the signal is emitted by a BNS (NSBH). The former is close to the
correct NS radius $R_{\rm \mbox{\tiny NS}}$, while the latter is biased.
The difference between the two depends on $R_{\rm \mbox{\tiny NS}}$ and the 
masses of the stars 
\begin{equation}\label{eq:Rdiff}
R_{\textrm{\mbox{\tiny BNS}}}-R_{\textrm{\mbox{\tiny NSBH}}}\equiv \Delta R(R_{\rm \mbox{\tiny NS}},m_{1},m_{2})>0,
\end{equation}
and it is plotted in Fig. 1 of~\cite{Yang:2017gfb}. 
The difference is smaller for larger $m_2$: the tidal deformability is a steeply decreasing function of the mass and almost negligible for a $2M_{\odot}$ NS. 
Misinterpreting a heavy BH for a NS induces almost 
negligible error in the radius estimate, but NSBHs with $1M_{\odot}$ BH result in a heavily biased radius estimate.

\subsection{Simulation of a population}

Now consider a population of $N$ low-mass binaries comprised mostly of BNSs, 
but possibly contaminated by a few 
NSBHs. Information from the BNSs will result in an unbiased estimate of the true NS radius $R_{\rm NS}$, 
while the corresponding radius estimate from the NSBHs will be biased by 
$\Delta R(R_{\rm \mbox{\tiny NS}},m_{1},m_{2})$.
To simulate such a population we assume that the signal-to-noise ratio 
(SNR), $\rho$, of each event follows the power-law 
distribution $3\rho_{\rm th}/\rho^4$~\citep{2011CQGra..28l5023S,2014arXiv1409.0522C}, 
where $\rho_{\rm th}\equiv 12$ is the network SNR detection threshold. 
This SNR distribution is a reasonable choice since the $(0.5-2)M_{\odot}$ detectable 
binaries will be relatively local 
(redshift less than 0.1) with current GW detectors~\citep{Aasi:2013wya}.
 
We draw NS and BH masses from a uniform distribution in $(0.5,2)M_{\odot}$
and set all NS radii to $R_{\rm \mbox{\tiny NS}}=12$km, consistent with the median radius 
measurement of~\cite{Abbott:2018exr}. 
The inferred radius for each event $i$ has a standard deviation $\sigma_{R_i}$ 
which is set to $\sim$0.75km at $\rho=33$, consistent with 
GW170817~\citep{Abbott:2018exr} and scales inversely
with the SNR of the event. 
The likelihoood for the inferred radius of each event is then approximated with a normal distribution centered at 
$R_i+{\cal{N}}(0,\sigma_{R_i})$ and with a standard deviation $\sigma_{R_i}$,
where $R_i=R_{\rm \mbox{\tiny NS}}$ if the event is a BNS, or $R_i=R_{\rm \mbox{\tiny NS}}-\Delta R(R_{\rm \mbox{\tiny NS}},m_{1i},m_{2i})$ 
if it is a NSBH. 
The additional scatter in the mean of the likelihood is caused by the random 
instance of detector noise. We approximate the likelihood for the component masses similarly, assuming 
a standard deviation of $\sigma_{m_i}=0.1 M_{\odot}$ at $\rho=33$~\citep{Abbott:2018exr}.

\subsection{Special Event Analysis}

Given the above population and corresponding radius measurements we first study whether 
we can determine the nature of individual events.
Our method is based on the fact that the inferred radii from the BNSs will be 
consistent with $R_{\rm NS}$, while the NSBHs result in a biased radius whose value depends on the 
component masses. 

We divide the $N$ detections into two groups: 
a special event whose type we want to determine and the remaining $N-1$ detections. 
We compute the Bayes Factor (BF) that the special event is a BNS compared to a NSBH
\begin{equation}\label{eq:bf}
{\rm BF}=\frac{\displaystyle\int p(R'|\mathscr{H}_{\rm \mbox{\tiny BNS}}) {\cal{L}}_s(d|R')\,dR'}{\displaystyle\int p(R'|\mathscr{H}_{\rm \mbox{\tiny NSBH}}) {\cal{L}}_s(d|R')\,dR'},
\end{equation} 
where ${\cal{L}}_s(d|R')$ is the radius likelihood for the special event given the GW data $d$ and 
$p(R'|\mathscr{H}_{\rm \mbox{\tiny BNS}})$ or $p(R'|\mathscr{H}_{\rm \mbox{\tiny NSBH}})$ is the prior assuming the event is a BNS 
or NSBH respectively. If ${\rm BF}>1(<1)$, the GW data are more consistent with the event being a BNS (NSBH). 

The radius likelihood for the special event is computed as detailed above, while the priors are computed 
by making use of the remaining $N-1$ events.
We multiply the radius likelihoods for the $N-1$ detections and obtain the combined 
likelihood $f(d|R)$. 
Assuming a low ratio of NSBHs to BNSs, or equivalently that the $N-1$ events are mostly
BNSs, $f(d|R)$ will be consistent with $R_{\rm \mbox{\tiny NS}}$.
Assuming a flat prior on the radius, the appropriately-normalized combined likelihood can be 
interpreted as the prior probability on
the radius for a BNS event not belonging in the $N-1$ detections, for example our special event: 
$p(R|\mathscr{H}_{\rm \mbox{\tiny BNS}})=f(d|R)$. If the special event is a NSBH, then the prior
can be computed again using $f(d|R)$ and shifting it by the expected radius bias
\begin{align}
p(R|\mathscr{H}_{\rm \mbox{\tiny NSBH}})=\displaystyle \int &p_s(m_1,m_2|d)\times \\ \nonumber
&f(d|R+\Delta R(R,m_1,m_2)) \, dm_1\,dm_2,
\end{align}
where $p_s(m_1,m_2|d)$ is the posterior of the two component masses of the special event. 

\begin{figure}[h]
\includegraphics[width=\columnwidth,clip=true]{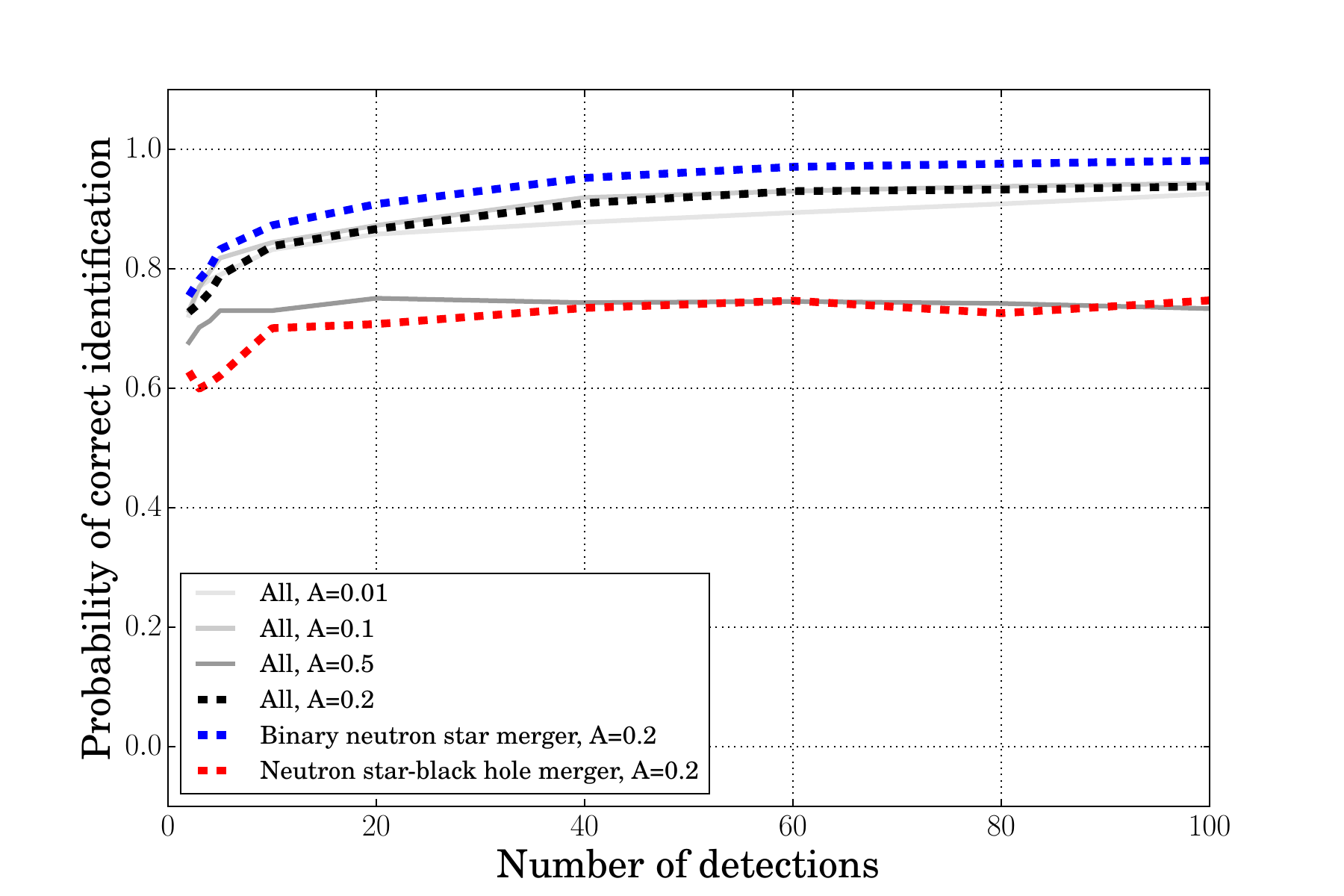}
\caption{Probability of correct identification of the highest-SNR event as a function of the number of detections.  
Thick dashed lines correspond to a rate ratio of NSBHs to BNSs of $A=20\%$.
The blue/red line is the probability of correct identification of a BNS/NSBH if the event is truly a BNS/NSBH.
The black line is the probability regardless of the event type. 
The grey lines are similar to the black line, but with 
the NSBH and BNS rate ratio of $1\%$, $10\%$, and $50\%$ (light to dark grey).
}
\label{fig:singleevent}
\end{figure}

We apply this method to simulated events.
We consider 1500 populations, compute the BF for each special event, and from those the probability
of correct identification. 
We find that we can correctly identify the binary type if the special event is selected wisely. 
In Fig.~\ref{fig:singleevent} we consider the highest-SNR event
as this event would have small uncertainty in radius and mass. 
We find that the highest-SNR event is 
correctly identified $80\%$ of the time after $\sim 5$ events
if $20\%$ of them are NSBHs. 
The overall probability of correct identification reaches $90\%$ after $\sim 40$ events. 
For larger ratios of NSBHs to BNSs, the NS radius prior might 
not represent the true radius. 
Such a biased measurement lowers the probability of correct identification. However, even if 
half the events are NSBHs, the probability of correctly classifying 
the highest-SNR event is $\sim 70\%$ after about 10 detections.  

\subsection{Hierarchical Mixture Model}

{The single event analysis allows for a high confidence identification/exclusion of NSBHs with a small number 
of events, however the analysis is only for identification purpose. In order to further 
measure the ratio of NSBHs to BNSs in a population and infer the NS radius we employ a hierarchical approach~\citep{2004AIPC..735..195L}}\cg{.}
The inferred radii follow a common underlying distribution which we model with a mixture model with two gaussian 
components and the likelihood
\begin{align}
{\cal{L}}\sim (1-A) {\cal{N}}(R_1,\alpha_1)+ A {\cal{N}}(R_2,\alpha_2).\label{hierL}
\end{align}
The first gaussian component models 
 the BNSs with a common radius $R_1$, while the second gaussian 
component models the NSBHs. We use a prior on $R_1$ that is uniform in $[10-14]$ km; for $R_2$ we use a uniform prior in 
$[R_1-10,R_1-3]$ since the inferred radii from NSBHs are smaller than the corresponding radii
from BNSs. The parameter $A$ is the
ratio of NSBHs to BNSs so we use a uniform prior in $[0,1]$. We assume that the rate ratio 
does not evolve with redshift, a reasonable assumption for low-mass binaries detected by second generation  detectors.

The scatter $\alpha_1$ in the radii of the BNSs
is caused by the detector noise realization. 
To find a suitable prior for $\alpha_1$
we analyze BNS-only populations and find that the posterior for $\alpha_1$ can be
approximated by a lognormal distribution with a mean of $0.8/\sqrt{N}$ km 
and a standard deviation of $1$ km. 
The scatter in the NSBH radii $\alpha_2$
is a combination of detector noise and the fact that the inferred radius from NSBHs depends on 
the bodies' masses. For lack of knowledge of the NSBH mass distribution we simply use a
wide prior for $\alpha_2$: a lognormal distribution with a mean of $\sqrt{3}$ km 
and a standard deviation of $1$ km. We have verified that all prior bounds do
 not affect the resulting posteriors, with the obvious exception of $A$.

\begin{figure}[h]
\includegraphics[width=\columnwidth,clip=true]{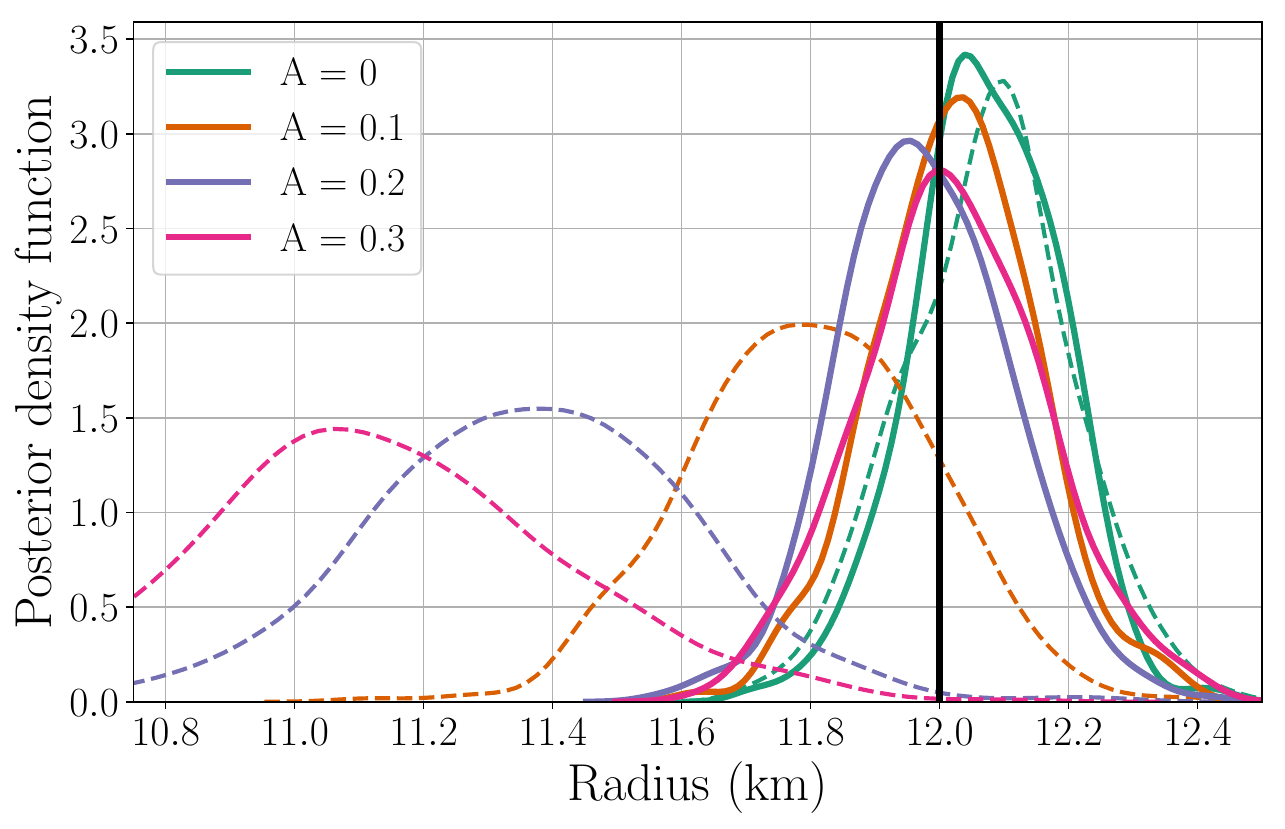}
\caption{Radius posterior density with the hierarchical mixture model (solid) lines for different ratios of NSBHs to BNSs and a population of 100 detections. In dashed we show the result of assuming the population contains only BNS, i.e. setting $A=0$ in Eq.~\eqref{hierL}. The vertical line is the true radius.}
\label{fig:Rmeasurement}
\end{figure}

We simulate populations of low-mass detections that are potentially contaminated by NSBHs 
and compute the posterior of the $5$ parameters of the hierarchical mixture model, Eq.~\eqref{hierL}.
This method can correct the bias in the NS radius estimate even
if the population includes NSBHs as we show in Fig.~\ref{fig:Rmeasurement} which plots the posterior
for $R_1$ with and without (setting $A=0$) the mixture model for different values of the ratio of NSBHs to BNSs. 
In all cases
the mixture model is able to separate the detected events well-enough into BNSs and outliers such 
that it leads to a correct estimate of the true BNS radius.

\begin{figure}[h]
\includegraphics[width=\columnwidth,clip=true]{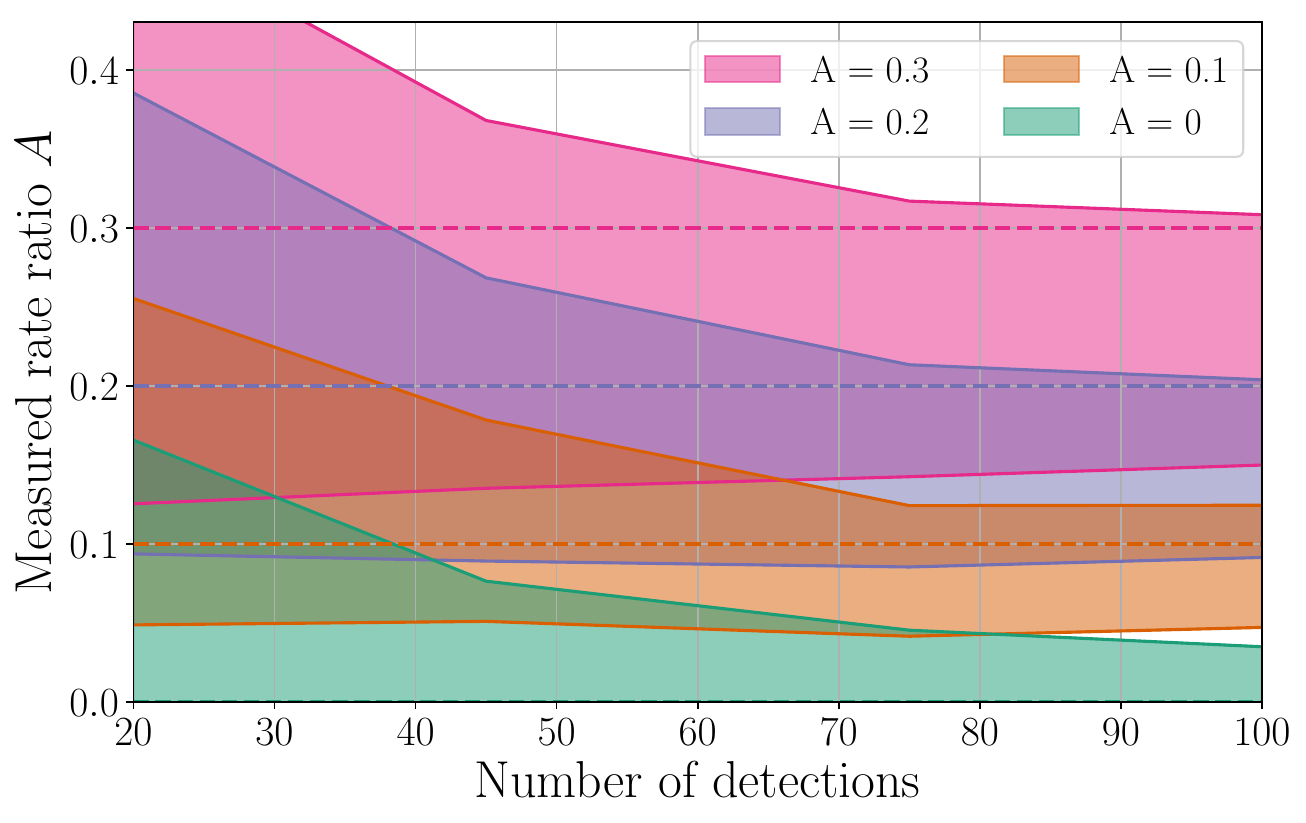}
\caption{Credible interval for $A$, the ratio of NSBHs to BNSs, as a function of the number of detections
 averaged over many populations for different simulated values of $A$. The green shaded region shows the
$90\%$ upper limit on $A$ for a BNS-only population, while the orange, blue, and pink regions show the 
$68\%$ credible interval for A when the ratio of NSBHs to BNSs is 0.1, 0.2, and 0.3 respectively (dashed horizontal lines).
}
\label{fig:A_posterior}
\end{figure}

Besides a corrected measurement of the NS radius, we also obtain an estimate
for $A$, the ratio of NSBHs to BNSs. In Fig.~\ref{fig:A_posterior} we plot credible
intervals for A as a function of the number of events, averaged over 200 populations. We find that if no low-mass NSBHs exist we put an upper limit on their relative abundance of $3\%$ at the $90\%$ level with 100 detections. If, on the other hand, low-mass NSBHs do exist we can constrain their abundance to within 0.16(0.11)[0.08] at the $68\%$ level with 100 detections if the true ratio is 0.3(0.2)[0.1].

\section{Discussion}

We present a method to identify NSBHs in a population of low-mass events, 
measure their relative abundance, and measure the NS radius.
We find that we can correctly classify the loudest events with only a handful of detections and 
measure the ratio of NSBHs to BNSs with a few dozens of events.
In fact, {the combined merger rate of GW170817 and GW190425 is $1090_{-800}^{+1720}$ Gpc$^{-3}$yr$^{-1}$~\citep{Abbott:2020uma}}. 
This applies to any merger in this mass range, be it a 
BNS or a NSBH, suggesting a few to many tens of relevant detections
in the upcoming observing runs~\citep{Aasi:2013wya}. 
We therefore expect the identification of a BNS or a NSBH with GWs alone in the near future
and a measurement of their rate ratio with a few years of data. 
We emphasize that we do not use information from the post-inspiral phase of the binary, or
rely on EM counterparts to the mergers. 

Our approach treats NSBHs as outliers in a population so its performance is degraded if the
fraction of NSBHs is high.
However, we show that the probability of correct identification of the event with the largest
SNR reaches $70\%$ after 
$10$ detections even if $50\%$ of the low-mass mergers are NSBHs. 
Similarly, we find that our ratio posteriors in Fig.~\ref{fig:A_posterior} are systematically shifted to lower 
values of $A$ as $A$ increases. Despite that, we can recover the rate ratio 
at the 1$\sigma$ level for a ratio up to at least $30\%$. {Moreover, we obtain an unbiased radius estimate
event for $A=30\%$ as our approach is based on identifying radius outliers; any potentially misidentified NSBH 
will have an inferred radius consistent with the BNSs and will thus not bias the radius estimate.}
	
For our simulations we assumed a true NS radius of $12$ km. A stiffer EoS, a heavier BH,
or a lighter NS will lead to a larger bias in the measured NS radius~\citep{Yang:2017gfb} and make
 classification {and measurement of the ratio $A$} easier. 
We also assume that the NS and BH masses are distributed uniformly in $(0.5,2)M_{\odot}$. 
If the NS mass distribution instead favors heavy stars while most BHs are lighter, 
both classification and the ratio measurement will improve. We expect the contrary if
low-mass BHs have masses around $2M_{\odot}$.
 
One caveat is that our analysis is formulated in terms of the NS radius
and the assumption that it is approximately constant for all BNSs, at least to within statistical errors.
This is reasonable for hadronic EoSs, but it is not expected to hold for EoSs with phase
transitions to quark matter~\citep{Han:2018mtj}. 
We do not consider this a limitation as our analysis can also be formulated
in terms of a quantity that it truly universal for all NSs: the EoS itself. 
In fact, the radius is correlated with the pressure at twice 
the nuclear density~\citep{Ozel:2016oaf}, suggesting that our arguments can be 
applied to the EoS directly. 
Specifically, a population of BNSs will yield an ever-improving 
measurement of the common EoS, while a misidentified NSBH will result in
 an EoS that is different than the population. 
 {In addition, the tidal deformability of neutron stars with exotic matter are likely to be similar to 
 stars with hadronic matter, so we do not expect a confusion between NSBHs and BNSs with exotic matter.}

Other systematic errors in the analysis might affect the inferred radius itself. 
We test this by artificially 
widening the radius likelihood by 500m, and find that 
the probability of correct identification is reduced by just $\sim 5\%$. 
In reality, Ref.~\cite{Abbott:2018wiz} argued that systematic errors are small even for a loud event
like GW170817 and they are likely to remain smaller than statistical errors until we detect a 
signal with an SNR of about 100~\citep{Dudi:2018jzn}.

As a final note, seeking outliers in a population can also be used to identify
exotic systems, such as binaries with at least one quark star, or hybrid binaries where one (or both)
stars have undergone a phase transition.

\section{Acknowledgements}
We acknowledge valuable discussions with Will Farr, Ramesh Narayan, Yu-Dai Tsai, Salvatore Vitale. 
We thank Reed Essick and Francesco Pannarale for useful comments.
H.-Y.C. was supported by by the Black Hole Initiative at Harvard University, which is funded by grants from the John Templeton Foundation and the Gordon and Betty Moore Foundation to Harvard University.
The Flatiron Institute is supported by the Simons Foundation.
Plots in this manuscript have been made with {\tt matplotlib}~\citep{Hunter:2007} and we have used {\tt stan}~\citep{JSSv076i01} to sample the mixture model.

\bibliography{OurRefs,moreref}

\end{document}